\documentclass[12pt]{JHEP3}
\newcommand{\BEQ}{\begin{equation}}
\newcommand{\EEQ}{\end{equation}}
\newcommand{\abs}[1]{\mbox{$\left| #1 \right|$}} 

\newcommand{\bra}[1]{\mbox{$\langle #1 \left| \right. $}}
\newcommand{\ket}[1]{\mbox{$ \left| \right.  #1 \rangle$}}

\title{Strings, Black Holes and the Extreme Energy Cosmic Rays}
\author{G.~Domokos
and S.~Kovesi-Domokos \\
 Department of Physics and Astronomy, The Johns Hopkins University, 
Baltimore, MD 21218, USA \\ E-mail:  \email{skd@jhu.edu}}

\abstract{In  a large class of 
string and string inspired  models the 
string excitation and black hole pictures invoked as an explanation of
trans-GZK cosmic ray events are equivalent. Single particle inclusive
distributions are asymptotically thermal at  the Hagedorn temperature. 
The hadron multiplicities are
reminiscent of multiplicities in heavy nucleus initiated interactions.}
\keywords{bhs, crp}
\preprint{JHU-TIPAC 2004}
\begin{document}

Primary cosmic rays of energies exceeding the Greisen-Zatsepin-Kuzmin (GZK) limit pose an exciting 
challenge to particle physics and/or astro\-physics. In essence, there are no known sources within 
about 50~Mpc capable of producing hadrons of energies substantially exceeding energies of 
$5\times 10^{19}$eV. Nevertheless, the trans-GZK cosmic rays produce air showers very similar 
in their structure to hadron induced ones. For a recent review of this puzzle, the reader may
consult Sigl's  article \cite{sigl1}. In order to resolve the puzzle, we conjectured~\cite{prl1} 
that the primaries are neutrinos, which are able to penetrate the cosmic microwave background 
(CMBR)
essentially uninhibited. In interactions with atmospheric nuclei, however, they have CMS
energies such that $\sqrt{s}$ exceeds the characteristic scale, $M_{\ast}$, of a low scale 
string theory \cite{lowscale}. Consequently, interactions are unified and string excitations produce a 
large cross section, thus generating hadron-like air showers. As an alternative,
Anchordoqui {\em et al.},  ref.~\cite{anchordocberg} proposed that ultra high energy neutrinos
produce ``mini black holes'' as originally suggested by Feng and Shapere \cite{blackholes} and
those are responsible for the trans-GZK cosmic ray events. The production cross section is 
assumed to be proportional to the Schwarzschild radius squared, {\em cf.} Dimopoulos and
Landsberg~\cite{dimolands}

In a more general context, the connection between string theories and black holes has been 
investigated over a number of years, following the original suggestion by Susskind~\cite{susskind}.
A careful study of the black hole scenario was  carried out by Ahn~{\em et al.} \cite{ahn}.
In general, those investigations fall into two different categories. Some authors
considered the statistical mechanics of gases made of strings,~\cite{skagerstam, bowick}.
Others, notably Amati and Russo~\cite{amatirusso} and Damour and Veneziano~\cite{damour}
performed explicit calculations within the framework of some specific string models 
and concluded that highly excited string states resemble (are identical to?) black holes.

In this article  we suggest  that the string excitation and mini black hole pictures of high energy
reactions are, in essence, equivalent, provided that the black holes are treated quantum
mechanically, as opposed to the semiclassical treatment given in 
refs.~\cite{anchordocberg, blackholes}. The conjectured equivalence holds
in a broad class of theories and we only need rather general features of them. This is fortunate:   
currently there is no really good candidate
for a phenomenologically viable string model. In fact, it has been repeatedly suggested that
string models are perhaps not directly relevant for  physics: one should rather think about
theories of a more general category, see, for instance, Johnson's book~\cite{cliff} for a very 
lucid exposition. Nevertheless, for the sake of definiteness, we use the term ``string'' in
what follows.
Furthermore, we exploit the conjectured equivalence of the two pictures in order 
to predict single particle inclusive distributions in the high energy reactions induced by 
trans-GZK cosmic rays: in principle, this prediction is testable by following the development of 
trans-GZK showers induced in the atmosphere. Here we concentrate mostly on qualitative aspects of the 
subject: a more detailed,  quantitative investigation is deferred to a future publication.

We illustrate our point on a simple example: the excitation of a high level string state. The
inclusive single particle distribution considered by Amati and Russo, {\em loc. cit.} can be dealt 
with similarly.
The levels of a string are labeled by a positive integer $N$ such that the mass of the state is 
given by the formula
\begin{equation}
M^{2} \sim  N M_{\ast}^{2} \qquad (N\gg 1)
\label{massquared}
\end{equation}
In eq.~(\ref{massquared}) $M_{\ast}$ stands for the characteristic string scale. 
In general, for $N\gg 1$ the state is a a highly degenerate one.  Let now ${\mathcal O}$
be an operator which creates a  string state of index $N$ from an initial state \ket{i}.
(We have in mind, for instance, string excitation in $\nu$-quark interactions, as conjectured in
ref.~\cite{prl1}. In that case, \ket{i} can stand for a quark state and ${\mathcal O}$ for the fermionic
current corresponding to the neutrino.)

Omitting trivial factors, the transition probability is given by:
\begin{equation}
P \propto \sum_{\alpha} \bra{i}{\mathcal O^{\dag}}\ket{N, \alpha}\bra{N, \alpha} {\mathcal O}\ket{i}\delta 
\left( E - NM_{\ast}\right)
\label{transitionprobability}
\end{equation}
In eq.~(\ref{transitionprobability}) $\alpha$ stands for the collection of labels necessary for the full
specification of states at level $N$: one has to sum over (most of) those, since the 
quantum numbers carried by \ket{i} and ${\mathcal O}$ do not fully specify the substate at level $N$
to which the transition takes place.

We  recognize that the quantity in eq.~(\ref{transitionprobability}),
\begin{equation}
\rho_{M}= \sum_{\alpha}\ket{N, \alpha}\bra{N, \alpha} \delta \left( E - NM_{\ast} \right)
\label{microcanonical}
\end{equation}
is just the (microcanonical) density matrix of the final state. (In statistical mechanics, a smoothing of
the delta function is necessary in order to define a level density; in the present context, 
such a smoothing is automatic if the finite width of resonances is taken into account.) The summation
over $\alpha$ leads to the information loss discussed by Amati~\cite{amati}.

It is well known that  a {\em large} microcanonical ensemble can be well approximated by a
canonical one, see, {\em e.g.} Huang's textbook~\cite{huang}. The argument given in textbooks 
runs as follows. One establishes that the fluctuations in a canonical ensemble  
asymptotically follow a normal distribution; the width of the distribution gets narrower 
as the energy  increases. In the limit of infinite energy the normalized distribution becomes 
a $\delta$ function.

A more direct approach is the following.
Let $E$ be the energy as before and $H$ be the Hamiltonian of the system. One can write 
the microcanonical density matrix in operator form as follows.
\begin{equation}
\rho_{M} = \delta \left( E - H \right)
\end{equation}

In the present context, the energy is the invariant center of mass energy of the interaction
described by eq.~(\ref{transitionprobability}), $E=\sqrt{s}$ and $H$ is the mass operator
of the string. 

One then uses the following complex integral representation of the $\delta$ function:
\begin{equation}
\delta \left( E - H \right) = \frac{1}{2\pi i}\int_{L- i \infty}^{L + i \infty} dv \exp  v\left(E - H\right),
\end{equation}
where an appropriate choice of $L$ gives a convergence factor. Next, one takes the trace of both sides
to obtain the partition functions. (This is sufficient: by adding external source terms to $H$,
one obtains the generating functional of the  correlation functions.)
One has then:
\begin{equation}
\exp S(E) = \frac{1}{2\pi i} \int_{L-i\infty}^{L+i\infty}dv \exp ( v E) \quad \exp \left( S_{c}(v)\right);
\end{equation}
We used the relation between the partition function and entropy on both sides of the previous equation.
Obviously, $\exp S_{c} (v)$ is the canonical partition function at inverse temperature $v$.
Next, one evaluates the integral in the previous equation by means of the method of steepest descents.
The location of the saddle point, $v=\beta$, is given by the equation:
\begin{equation}
E = - \left(\frac{\partial S_{c}}{\partial v}\right)_{v=\beta},
\end{equation}
which is the usual thermodynamic relation between temperature and energy.
Using the preceding equation, the  leading order contribution of the integral is given by 
\begin{equation}
\exp (S_{c}(\beta) - E\beta) = \exp S(E).
\end{equation}
The quantity $S_{c} \beta - E\beta$ is recognized as the Legendre transform of the canonical entropy.
Finally, by carrying out the Gaussian integration, one obtains a contribution to the
entropy, which, up to an irrelevant additive constant equals to $ -3/2 \ln E/M_{\ast}$.
All these formulae are valid for $E/M_{\ast} \gg 1$.

An
operational definition of a ``large'' microcanonical ensemble is
that terms of $O(1/E)$  in the expression of the entropy  are negligible\footnote{Amati and 
Russo {\em loc. cit.} use an adaptation of the Darwin-Fowler method and discuss some 
terms of $O(1/N)$ in the framework of specific string models. For $N\gg 1$, the asymptotic
approach used here and the Darwin-Fowler method yield identical results.}.

Here we consider level densities of the form:
\begin{equation}
d(E)=\exp S(E)  \sim  C \left( E/M_{\ast}\right)^{-\gamma -3/2}\exp \left(\alpha (E/M_{\ast})^{\delta}\right); 
\end{equation}

The quantities $ C, \gamma,  \alpha,  \delta $ depend on the specific model 
considered. The contribution of the Gaussian integral to the entropy has been
taken into account. (In known string models, $\delta = 1$.)
The inverse temperature is:
\begin{equation}
\beta = \frac{\partial S}{\partial E} \sim  - \frac{\gamma +3/2}{E} + \alpha \delta\left(\frac{E}{M_{\ast}}\right)^{\delta -1}M_{\ast}^{-1}
\label{inversetemperature}
\end{equation}
Notice that for $\delta = 1$, the temperature is asymptotically constant and, hence, it may be
identified with a Hagedorn temperature, $T_{H}$. Notice, however that the expression of the inverse 
temperature, eq.~(\ref{inversetemperature}) gives a temperature approaching $T_{H}$ {\em from above} From a physical point of view, this is unacceptable: 
the Hagedorn temperature is supposed to be a {\em maximal} temperature.
This is due to the fact that the coefficient of $1/E$ in eq.~(\ref{inversetemperature}) is
negative. The problem is corrected by taking into account the contribution of Kaluza-Klein
excitations to the entropy. Apart from an additive constant, the contribution
is proportional to $(E/M_{\ast})^{n}$, n being the number of compact dimensions.
For instance, for an open superstring in the critical dimensionality, $n=6$ and $\gamma = 9/2$, giving
$-(\gamma + 3/2) + n =0$. Thus, the approach to the Hagedorn temperature in this case is 
determined by higher order terms in the asymptotic expansion. 

 The expression of 
the Hagedorn temperature 
in terms of $M_{\ast}$ is somewhat model dependent: in the calculations cited above, one has,
$T_{H}= M_{\ast}/(a \pi)$, with $a$ ranging, approximately, between 2 and 4 for various string models. 
For the purposes of
numerical estimates in this work, we adopt $a=3$: this is equivalent, of course,  to fixing 
the parameter $\alpha$ in eq.~(\ref{inversetemperature}).
Equation~(\ref{inversetemperature}) establishes the connection between the string and (microscopic)
black hole pictures: in fact, it means that -- in a sense -- a large and highly degenerate isolated 
system serves as
its own thermal reservoir. Consequently, the terms ``string'' and ``black hole'' are used interchangeably
in what follows.

The asymptotic estimate given here has far reaching consequences. In particular, the inclusive single
particle distribution can be calculated using  elementary statistical mechanics
of a canonical ensemble as it was carried out in detail for specific string models by
Amati and Russo, ref.~\cite{amatirusso}. 
We have in the  rest frame of a decaying resonance:
\begin{equation}
dn = \frac{d\sigma}{\sigma}= \frac{1}{\exp \left(k_{0}/T\right)\pm 1} \frac{d^{3}k}{k_{0}}L^{2},
\label{restframe}
\end{equation}
where $k_{\mu}$ stands for the four momentum of the particle observed in the decay and,
asymptotically, we may put $T=T_{H}$. The sign $\pm$ 
is valid for Fermi/Bose statistics of the observed particle, respectively.
The quantity $L$ is a characteristic length of the system. Tentatively we put
it equal to a compactification radius in a symmetric toroidal compactification scheme. In what follows,
we also  put $k^{2}=0$. This is a permissible approximation except perhaps in the far infrared of the
observed particle which, however, is not very relevant for the development of observable showers.

We now note that the expression of $dn$ is Lorentz invariant. Consequently, one can replace the 
rest frame expression of $k_{0}$ by a Lorentz invariant quantity reducing to $k_{0}$ in that frame.
This is elementary. Use the invariant:
\begin{equation}
K= \frac{\left(P\cdot k\right)}{\sqrt{P^{2}}}
\label{invariantenergy}
\end{equation}
in place of $k_{0}$ in eq.~(\ref{restframe}). Here, $P$ stands for the total four momentum of the 
decaying resonance. In a scattering process, such as 
\[ \nu +  q \longrightarrow \mbox{resonance} \longrightarrow (k) + X, \]
one can approximate $\sqrt{P^{2}}\approx \sqrt{s}$. (Here, ``$(k)$'' stands for the observed
particle of four momentum $k$.)
The inclusive distribution, eq.~(\ref{restframe}) with the replacement $k_{0}\longrightarrow K$
shows the expected properties if evaluated in the laboratory frame.
In the high energy limit, fermions and bosons have to be treated somewhat differently, due to
their different statistics.

We evaluate the invariant $K$ in the laboratory frame, where the target quark of an effective
mass $m$ is approximately at rest and the incident neutrino has energy $E_{\nu}$.
In the limit $E_{\nu}\gg m$ we have:
\begin{equation}
K \sim k_{0, L}\sqrt{\frac{E_{\nu}}{2m}}\left( \frac{m}{E_{\nu}} + 2 \sin^{2}\left(\theta /2\right) \right).
\label{Kinlabframe}
\end{equation}
 Here, $k_{0,L}$ stands for the energy of the observed particle evaluated in the laboratory frame;
$\theta$ is the angle between the incident neutrino and the observed particle.

There is a forward peak ($\theta =0$) both for a boson and a fermion observed. For a boson, we have
\[ \frac{1}{\sigma}\frac{k_{0}d\sigma}{d^{3}k}\left|_{\theta = 0}\sim   
\frac{T_{H}}{k_{0,L}}\sqrt{\frac{2E_{\nu}}{m}}L^{2},\right. \]
whereas for a fermion the term $\propto m/E_{\nu}$ can be neglected in eq.~(\ref{Kinlabframe}) and 
one finds that the forward peak remains finite:
\[ \frac{1}{\sigma}\frac{k_{0}d\sigma}{d^{3}k}\left|_{\theta = 0}\sim \frac{L^{2}}{2}.\right. \]
The different behavior of observed bosons and fermions is due, of course, to their different 
statistics. Obviously, $dn$ is exponentially small for 
a finite value  of the emission angle $\theta$, as expected: this is just the Boltzmann limit
of the thermal distribution.

One can evaluate the total multiplicities by means of a straight forward  integration. 
As stated before,  neglecting the rest masses of the observed particles is a fair
approximation. With that, one has:
\begin{equation}
I \equiv  L^{2} \int \frac{d^{3}k}{k_{0}} \frac{1}{\exp (k_{0}/T_{H}) -1} = \frac{L^{2}}{12}T_{H}^{2} ,
\label{bosonintegral}
\end{equation}
 The corresponding integral for
fermions is smaller  by a factor of $1/2$.

It follows that the total multiplicity of a particle of any kind is proportional
to its statistical weight, {\em viz.}
\begin{equation}
n_{B} =\frac{1}{12} L^{2}T_{H}^{2} g_{B},
\label{bosonmultiplicity}
\end{equation}
and
\begin{equation}
n_{F} =\frac{1}{24}  L^{2}T_{H}^{2} g_{F},
\label{fermionmultiplicity}
\end{equation}
respectively. {\em In the asymptotic limit the multiplicities are
independent of the incident energy: an observable property of this scheme.}

As far as {\em unconfined particles} ($\gamma$, W, Z, leptons) are concerned,
eqs.~\mbox{(\ref{bosonmultiplicity}, \ref{fermionmultiplicity})} contain the final answer. 
However,
for quarks and gluons, one has to consider their fragmentation into observable hadrons.
Instead of carrying out a detailed analysis of the fragmentation, we adopt here an approximate
procedure, which, however, should give us an idea about the order of magnitude of
hadrons of any flavor produced.

We notice that popular analytical fits to parton fragmentation functions, on the average, 
give about $\langle n_{H} \rangle \approx 3 $ hadrons produced from each parton,
see \cite{bargerphillips} for a typical reference. Consequently, in order to obtain the
multiplicities of hadrons produced, we just multiply the quark and gluon multiplicities by $3$.

Below, we list  the statistical weights for each kind of particle considered, taking three
families into account for quarks and leptons.
\begin{eqnarray}
g_{q} + g_{\bar{q}} & = & 72 \nonumber \\
g_{g}& = & 16 \nonumber \\
g_{l} + g_{\bar{l}} & = & 24 \nonumber \\
g_{W} + g_{\bar{W}} + g_{Z} + g_{\gamma} & = & 8
\label{statweights}
\end{eqnarray}
In compiling the statistical weights, we assumed that in the string regime symmetries are unbroken; 
in particular, $W$ and $Z$ are transverse. For the time being, we did not consider the various 
superpartners which may be produced in such reactions. 

Finally, we have to determine the characteristic length, $L$, occurring in eq.~(\ref{restframe}).
We use the ADD formula~\cite{ADD}, assuming, for the sake of simplicity a symmetric toroidal
compactification:
\begin{equation}
\frac{1}{L}= 2\pi M_{\ast} \left( \frac{M_{\ast}}{M_{P}}\right)^{2/n}
\label{ADD}
\end{equation}
In eq.~(\ref{ADD}), $M_{P}$ stands for the four dimensional Planck mass and $n$ is the number of 
compactified dimensions. Previously we presented evidence that $M_{\ast}$ should be around
80~TeV or so,~\cite{amsterdam02}. In what follows, we use that value of $M_{\ast}$ for the
purposes of numerical estimates. However,
given the uncertainties at the present stage, it is reasonable to conjecture
$50{\mbox TeV} \leq M_{\ast} \leq 100{\mbox TeV}$. 

Using these ingredients, we  estimate the total number of hadrons generated in the initial
interaction. One has the estimate:
\begin{equation}
N_{had}\approx 3\times \left( g_{g} + 1/2 \left(g_{q} + g_{\bar{q}}\right)\right)I
\label{hadronmultiplicity}
\end{equation}

It is instructive to compare the multiplicity obtained from eq.~(\ref{hadronmultiplicity}) at a
reasonable number of extra dimensions ($n=6,7,8$) with what an incident proton or nucleus would 
produce\footnote{$n=6,7$ are theoretically motivated: superstrings and $n=11$ SUGRA, respectively.
We added $n=8$ in order to keep possible future models in mind. Low numbers of extra dimensions are 
probably excluded, see \cite{raffelt}}. In Table~1 we summarize the hadron
multiplicities estimated  in the black hole picture, using $M_{\ast}=80$TeV.
\TABLE{
\begin{tabular}{|c||c|c|c|}\hline
n & 6 & 7 & 8\\ \hline
$N_{had}$ & $ 1.76\times 10^{4}$ & $4.16\times 10^{3}$& $1.12\times 10^{3}$ \\ \hline
\end{tabular}
%
\caption{Hadron multiplicity for various numbers of extra dimensions}}
The number of prompt leptons can be calculated analogously.

 Let us recall that according to available data, the trans-GZK air showers
are ``hadron-like'': the showers develop similarly to hadron induced ones at lower energies,
see Sigl,~{\em loc. cit.}. Let us consider a primary energy, $E_{L}=3\times 10^{11}$GeV, about
the highest energy observed,~\cite{sigl1}. Assuming no new physics between present accelerator 
energies  and that energy, one can extrapolate the multiplicity of secondary hadrons in 
various reactions ($pp, p\bar{p}, e\bar{e}$). The energy dependence of the multiplicity 
is similar in all those reactions, {\em cf.} the tables in the Particle Data Group,~\cite{pdg}.
Lumping all  the data together, one obtains a simple fit:
\begin{equation}
N_{ch} = 2.60 - 0.18 f + 1.23 f^{2},
\label{multiplicityfit}
\end{equation}
where $f = s/1\mbox{GeV}^{2}$. Extrapolated to $E_{L}= 10^{11}$GeV in a $pp$ interaction,
one gets $N_{ch}\approx 170$. The total multiplicity, relevant for shower development,
is estimated by multiplying this number by 3/2, since most of the secondaries are pions.

Clearly, the hadronic multiplicities produced by the microscopic
 black holes are  
larger; thus, one may conclude that the event in question was 
initiated by a heavy nucleus.
Adopting a simple superposition model of nuclear interactions, see {\em e.g.} 
Gaisser~\cite{gaisser} one can estimate an  ``effective atomic number'' of the reaction by
taking the ratio of the multiplicity produced by a black hole to that of the extrapolated 
hadronic multiplicity.
All such formulae  are based on a number of assumptions and cannot be
used for obtaining quantitative estimates. 

In the following Table we summarize the effective atomic numbers, $A_{\mbox{eff}}$,
for the number of compactified dimensions considered here.

\TABLE{
\begin{tabular}{|c|c|c|c|}\hline
n & 6 & 7 & 8\\ \hline
$A_{\mbox{eff}}$  & 69  & 16 & 4 \\ \hline
\end{tabular}
\caption{Effective atomic number as a function of n}}

Given the crudeness of the approximations used in obtaining the entries of 
the preceding Table, this is
merely a hint to be followed up by more careful calculations. The analysis of future data from
detectors presently under construction will, eventually, bear out the behavior of showers induced by
trans-GZK cosmic rays. 

We conclude with a few remarks.
\begin{itemize}
\item We demonstrated the equivalence of a model of neutrino induced string excitations and one involving
(microscopic) black holes, both purporting to explain trans-GZK cosmic rays. The equivalence simplifies
the calculation of single particle inclusive distributions within such a scheme. In fact, at least 
asymptotically, the inclusive distributions can be calculated in a simple way, using Lorentz invariance 
and elementary statistical mechanics. It is to be pointed out that the asymptotic equivalence of the 
``excited string''  and ``mini black hole'' pictures is independent of the tree approximation used by Amati
and Russo, {\em loc. cit.}. It is, nevertheless, pleasing that an explicit calculation bears out the
the black body nature of a highly excited string.
\item While we concentrated on qualitative aspects of our discussion, it is likely that this can 
be followed up by a more quantitative study, involving a MC simulation of the string/black hole
induced trans-GZK showers.
\item There remain some  questions to be clarified. In particular,
the leading asymptotic estimates are largely model independent, but one has little knowledge
about {\em model dependent} corrections of $O(1/E)$. Very often, however, leading terms in an 
asymptotic expansion yield quantitatively correct estimates\footnote{A famous example is
the evaluation of $\Gamma (z)$ by means of the Stirling formula even for $\abs{z}$ not much 
larger than 1.}.
\end{itemize}

The idea of the  research reported here arose in discussions with  Zolt\'{a}n Kunszt
during the authors' visit at ETH in Zurich: we are grateful to Professor Kunszt for the
hospitality extended to us and a fruitful exchange of ideas. We further thank
Luis Anchordoqui, Will Burgett  and Haim Goldberg for several enjoyable discussions on the puzzle of trans-GZK
cosmic  rays.


\begin{thebibliography}{99}
\bibitem{sigl1} G. Sigl, Annals of Physics {\bf 303}, 117 (2003).
\bibitem{prl1} G. Domokos and S. Kovesi-Domokos, Phys. Rev. Lett. {\bf 82}, 1366 (1999).
\bibitem{lowscale} P.~Horava and E.~Witten, Nuc.~Phys. {\bf B475}, 94 (1996)
J.~Lykken, Phys.Rev. {\bf D54}, 3693 (1996). 
N.~Arkani-Hamed, S.~Dimopoulos and G.R.~Dvali, Phys.Lett. {\bf B429}, 263 (1998).
I.~Antoniadis, N.~Arkani-Hamed, S.~Dimopoulos and G.R.~Dvali, Phys. Lett. {\bf B4336}, 257 (1998).
\bibitem{anchordocberg} L.A.  Anchordoqui and Haim Goldberg, Phys. Rev. {\bf D65}, 047502 (2002).
L.A.~Anchordoqui, J.L.~Feng, H.~Goldberg and A.~Shapere, Phys.Rev. {\bf D65}, 124027 (2002).
\bibitem{blackholes} J.L. Feng and  A.D. Shapere, Phys. Rev. Lett. {\bf 88} 021303 (2002).
\bibitem{dimolands} S.~Dimopoulos and G.~Landsberg, Phys. Rev. Lett. {\bf 87} 161602 (2001).
\bibitem{susskind} L.~Susskind, hep-th/9309145 (unpublished).
\bibitem{ahn} E.-J. Ahn, M.~Ave, M.~Cavaglia and A.V.~Olinto, Phys. Rev. {\bf D68}, 043004 (2003).
\bibitem{skagerstam} P.~Salomonson and B.-S.~Skagerstam, Nuclear Physics {\bf B268}, 
349 (1986.)
\bibitem{bowick} M.J.~Bowick and L.C.R.~Wijewardhana, Phys.Rev. Letters {\bf 54}, 2485 (1985).
\bibitem{amatirusso} D. Amati and J.G.~Russo, Phys.~Letters {\bf 454}, 207 (1999).
\bibitem{damour} Th. Damour and G. Veneziano, Nuc. Phys. {\bf B568}, 93 (2000).
\bibitem{cliff} Clifford V. Johnson, ``D-Branes''. Cambridge Monographs on Mathematical Physics,
Cambridge University Press (2003).
\bibitem{amati} D.~Amati, in Proceedings of The Abdus Salam Memorial Meeting, edited 
J.~Ellis, F.~Hussain, T.~Kibble, G.~Thompson and M.~Virasoro. World Scientific Publishing Co.
Singapore (1997).
\bibitem{huang} K.~Huang, ``Statistical Mechanics'', second edition. Wiley \& Sons, 1987.  Secs.~6.2~and~8.3)
\bibitem{hornzachariasen} D.~Horn and F.~Zachariasen, ``Hadron Physics at Very High Energies'',
Ch.~6. Benjamin, New York (1973). References to the original articles are given in this work.)
\bibitem{bargerphillips} V.D.~Barger and R.J.N.~Phillips, ``Collider Physics'', Ch.~6.
Addison-Wesley (1987).
\bibitem{ADD} N.~Arkani-Hamed, S.~Dimopoulos and G.~Dvali, Phys.~Lett. {\bf B429}, 263 (1998).
\bibitem{amsterdam02} W.S.~Burgett, G.~Domokos and S.~Kovesi-Domokos, hep-ph/0209162.
(This is an expanded version of a paper presented at ICHEP~2002, Amsterdam.)
\bibitem{raffelt} S.~Hannestad and G.G.~Raffelt, Phys.~Rev.~Lett. {\bf 88}:071301 (2002).
\bibitem{pdg} K.~Hagiwara {\em et al.}, Phys.Rev. {\bf D66}, 010001 (2002). (http://pdg.lbl.gov)
\bibitem{gaisser} T.K.~Gaisser, ``Cosmic Rays and Particle Physics'', Ch.~16.
Cambridge University Press, Cambridge (1990).
\bibitem{bird} D.J.~Bird {\em et al.}, Astrophys. J. {\bf 441}, 144 (1995)
\end{thebibliography}
\end{document}